\documentclass{article}
\usepackage{spconf,amsmath,graphicx,amssymb}
\usepackage[belowskip=2pt,aboveskip=6pt]{caption}

\DeclareMathAlphabet{\mathpzc}{OT1}{pzc}{m}{it}

\newcommand{\argmin}{\mathop{\rm arg~min}\limits}

\graphicspath{{Images/}}

\title{An Energy-Efficient Compressive Sensing Framework Incorporating Online Dictionary Learning for Long-term Wireless Health Monitoring}
\name{Kai XU, Yixing Li, Fengbo Ren}

\address{Parallel Systems and Computing Laboratory (PSCLab) \\
		 Arizona State University}

\begin{document}

\maketitle

\begin{abstract}
Wireless body area network (WBAN) is emerging in the mobile healthcare area to replace the traditional wire-connected monitoring devices. As wireless data transmission dominates power cost of sensor nodes, it is beneficial to reduce the data size without much information loss. Compressive sensing (CS) is a perfect candidate to achieve this goal compared to existing compression techniques. In this paper, we proposed a general framework that utilize CS and online dictionary learning (ODL) together. The learned dictionary carries individual characteristics of the original signal, under which the signal has an even sparser representation compared to pre-determined dictionaries. As a consequence, the compression ratio is effectively improved by 2-4x comparing to prior works. Besides, the proposed framework offloads pre-processing from sensor nodes to the server node prior to dictionary learning, providing further reduction in hardware costs. As it is data driven, the proposed framework has the potential to be used with a wide range of physiological signals.
\end{abstract}


\section{Introduction}
The existing heathcare model of the medical system is based on episodic examination or short-term monitoring for disease diagnosis and treatment. The major issues in such a system are the overlook of individual variability and the lack of personal baseline data, due to limited frequency of measurements. Continuous or non-intermittent monitoring is the key to create big data of individual health record for studying the variability and obtaining the personal baseline. Recent advancements in wireless body area networks (WBAN) and bio-sensing techniques has enabled the emergence of miniaturized, non-invasive, cost-effective wireless sensor nodes (WSNs) that can be deployed on the human body for personal health and clinical monitoring \cite{5771056}.  Through WBAN, the monitored data can be transmitted to a near-field mobile aggregator for on-site processing. Through Internet infrastructures, the data can be uploaded to remote servers for storage and data analysis. These technology advancements will eventually transform the existing model of health related services to continuous monitoring for disease prediction and prevention \cite{Varshney}. Such a wireless health revolution will make healthcare systems more effective and economic, benefiting billions of individuals and the society they live in.

One of the key challenges faced by the long-term wireless health monitoring is the energy efficiency of sensing and information transfer. Due to the limited battery capacity of WSNs, continuous sensing inevitably increases the frequency of battery recharging or replacement, making it less convenient for practical usage. In the WSNs for bio-sensing applications, the energy cost of wireless transmission is about 2 orders of magnitude greater than other operations (e.g., analog-to-digital conversion (ADC)). State-of-the-art radio transmitters exhibit energy efficiency in the nJ/bit range while every other component consumes at most tens of pJ/bit \cite{6155205}. Therefore, reducing the data size for information transfer is the key to improve energy efficiency.

The CS framework \cite{5771056, islped_opt} offers a universal and simple data encoding scheme that can compress a variety of physiological signals, providing a viable solution to realizing energy-efficient WSNs for long-term wireless health monitoring. However, the compression ratio (CR) demonstrated by existing frameworks is limited given a signal recovery quality required for diagonosis purposes. In \cite{6313763,6637862} percent root-mean-square difference (PRD) of 8.5$\%$ and 9$\%$ is reported at a CR of 5x and 2.5x for ECG signals, respectively. These frameworks all deal with the sparsity of physiological signals on pre-determined bases and fail to take into account the individual variability in signals that is critical to exact signal recovery. 

In this paper, we propose an energy-efficient data acquisition framework, customized for the long-term electrocardiogram (ECG) monitoring, which exploits online dictionary learning (ODL) on server nodes to train personalized bases that capture the individual variability for further improving the sparsity of ECG signals. By incorporating such prior knowledge into signal recovery, the CS performance in terms of accuracy-CR trade-off is significantly enhanced, leading to further data size reduction and energy saving on sensor nodes. Additionally, the proposed framework does not require any pre-processing stages on sensor nodes. Alternatively, high reconstruction quality is enforced by pre-processing training data prior to the dictionary learning stage, to eliminate the impact of noise and interference on trained bases, enabling simpler and more cost-effective sensor structures. Experimental results based on MIT-BIH database show that  our framework is able to achieve an average PRD of $9\%$ at a CR of 10x. This indicates that our framework can achieve 2-4x additional energy saving on sensor nodes (for the same reconstruction quality) compared to the reference designs \cite{5771056, 6313763, 6637862, 6346966}. Due to the training and personalization of the dictionary, the proposed framework has the potential to be generally applied to a wide range of physiological signals.  

\section{Background}

\subsection{Compressive Sensing}
Assuming a signal $\mathbf{f} \in \mathbb{R}^{n}$ can be well represented by a sparse vector $\mathbf{x} \in \mathbb{R}^{k}$ on a certain basis $\mathbf{\Psi} \in \mathbb{R}^{n \times k}$ as $\mathbf{f}=\mathbf{\Psi} \mathbf{x}$, then the signal information can be well preserved by projecting $\mathbf{f}$ onto a random domain through a sensing matrix $\mathbf{\Phi} \in \mathbb{R}^{m \times n}$ (m$<$n) \cite{CandesIntroCS}, given as 
\begin{equation}
	\mathbf{y}=\mathbf{\Phi} \mathbf{f}=\mathbf{\Phi} \mathbf{\Psi} \mathbf{x}.
\end{equation}

Candes and et al. \cite{1580791} has proven that one has a high probability to recover the sparse coefficient $\mathbf{x}$ by solving the basis pursuit (BP) problem defined as
\begin{equation}
\min_{\mathbf{x}\in \mathbb{R}^{k}}{\Vert{\mathbf{x}} \Vert_{1}} \quad s.t.\quad \Vert \mathbf{y-\Phi \Psi x} \Vert_{2} \leq \varepsilon,
\end{equation} 
where $\varepsilon$ is an error tolerance term for enhancing the accuracy of the solution considering noise.

\subsection{Dictionary Learning}

Learning dictionaries from data instead of using off-the-shelf bases has been proved effective in improving signal reconstruction performance for images \cite{4011956}. The most recent dictionary learning algorithms \cite{aharon, olshausen,NIPS2006_2979} are second-order iterative batch procedures that access the whole training set at each iteration in order to minimize a cost function under certain constraints. Although these algorithms \cite{aharon, olshausen,NIPS2006_2979} have been shown experimentally faster than first-order gradient descent methods, they cannot effectively handle very large training sets \cite{Bottou08thetradeoffs}, because of the involved matrix factorization upon the entire training data. To be able to deal with large data sets for long-term monitoring, the ODL algorithm is adopted in our framework. Compared to the methods mentioned above, ODL has a higher training speed and requires less storage space \cite{mairalC} because of its elimination of large matrix factorizations. With ODL, it is possible to add new features into the dictionary without stalling the reconstruction, which offers a mechanic of amelioration when a distinctive input is received.

\subsection{ODL}

Assuming the training set is composed of i.i.d. samples following a distribution $p(x)$, ODL draws one sample $x_{t}$ at a time and alternates between the sparse coding stage and dictionary update stage.

\subsubsection{Sparse Coding}
The sparse coding problem is a $\mathit{l}_{1}$-regularized least-squares problem defined as
\begin{equation}
\mathbf{\alpha}_{t}=\argmin_{\alpha \in \mathbb{R}^{n}} \frac{1}{2} \Vert \mathbf{x}_{t}-\mathbf{D}_{t-1} \mathbf{\alpha} \Vert_{2}^{2}+\lambda \Vert \mathbf{\alpha} \Vert_{1}.
\end{equation}

Due to the high correlations between columns of the dictionary, a Cholesky-based implementation of the LARS-Lasso algorithm, which provides the whole regularization path, is chosen here to solve the sparse coding problem \cite{mairalJ}. 

\subsubsection{Dictionary Updating}
At this stage, the objective is to find a dictionary $\mathbf{D}$ that satisfies:
\begin{equation}
\mathbf{D}_{t} = \argmin_{\mathbf{D}} \frac{1}{t} \sum_{i=1}^{t} \frac{1}{2} \Vert \mathbf{x}_{i}-\mathbf{D} \mathbf{\alpha}_{i} \Vert_{2}^{2} + \lambda \Vert \mathbf{\alpha}_{i} \Vert_{1}.
\end{equation}

The problem in (4) can be solved by the block coordinate descent algorithm \cite{mairalJ}. Overall, the detailed procedure for ODL algorithm is summarized in Algorithm 1.
\vspace{2mm}
\hrule
\vspace{1mm}
\noindent \textbf{Algorithm 1} Pseudocode for ODL
\vspace{1mm}
\hrule
\vspace{2mm}
\noindent Input: Input data $x\in \mathbb{R}^{n}\sim p(x)$, initial dictionary $\mathbf{D_{0}}\in \mathbb{R}^{n \times k}$, number of iterations t.

\noindent Output: Learned dictionary $\mathbf{D}_{t}$.

\noindent Steps: 

1: Set $\mathbf{A}_{0}\leftarrow 0, \mathbf{B}_{0}\leftarrow 0$.

2: For t=1:T

3: \quad Draw a new sample $\mathbf{x}_{t}$ from $p(x)$.

4: \quad Sparse coding: find a sparse coefficient of $\mathbf{x}_{t}$ under current dictionary $\mathbf{D}_{t-1}$.

5: \quad  $\mathbf{A}_{t} \leftarrow \mathbf{A}_{t-1}+\alpha_{t}\alpha^{T}_{t}$.

6: \quad $\mathbf{B}_{t} \leftarrow \mathbf{B}_{t-1}+\mathbf{x}_{t}\alpha^{T}_{t}$.

7: \quad Dictionary update: update dictionary $\mathbf{D}_{t-1}$ column by column, the j-th column is given by

8: \quad For j=1:k

9: \quad \quad $\mathbf{d}_{j}\leftarrow \frac{1}{(A_{t})_{jj}}(\mathbf{B}_{t}(:,j)-\mathbf{D}\mathbf{A}_{t}(:,j))+\mathbf{d}_{j}$. 

	\quad \quad \quad if $\Vert \mathbf{d}_{j} \Vert_{2}>1$, then normalize it to unit form.

10: \quad end for

11. end for

12. Return $\mathbf{D}_{t}$.

\noindent\rule{9cm}{0.4pt}
\begin{figure*}[!bt]
	\centering
	\includegraphics[width=\textwidth]{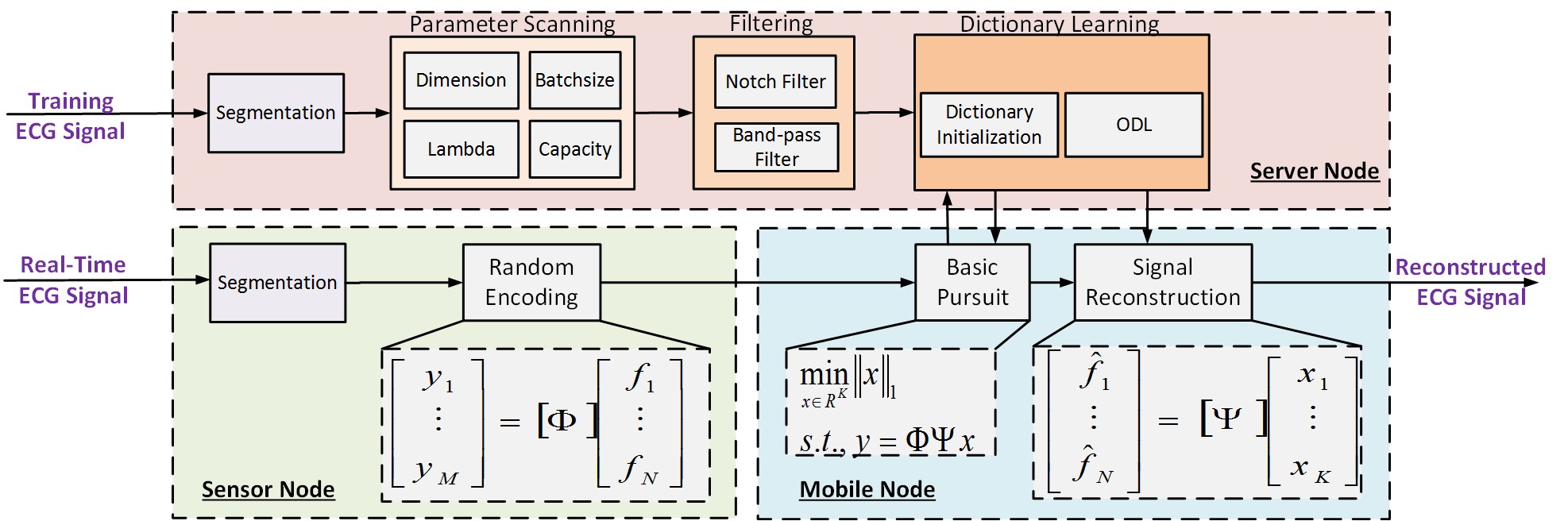}
	\caption{Block diagram of the proposed framework. The parameter sweeping and dictionary training procedure are executed on servers. The reconstruction process is performed on mobile platform for providing timely feedback. The random encoding process using random Bernoulli matrix is embedded into the sensor node for effective data compression and energy saving.}
\end{figure*}

\section{The Proposed Framework}

The most recent frameworks on ECG monitoring \cite{7015915, 5946515, Abo} adopt a QRS detection process, such as the Pan-Tompkins algorithm, prior to the sensing stage in order to locate the period information of ECG signals. However, integrating the QRS detection process into the sensor nodes not only occupying CPU cycles but also burning excessive power. For wearable applications, an energy-efficient framework must get rid of such pre-processing stages on sensor nodes.

The block diagram of the proposed framework is shown in Fig. 1. It is composed of three functional modules (i.e., dictionary learning, random encoding, and CS signal reconstruction, performed on a server node, a sensor node, and a mobile node, respectively).

The dictionary learning module is used to train personalized bases to capture the individual variability that is critical to exact signal recovery. As dictionary learning directly extracts features from the segmented raw data, the learned dictionary contains critical temporal and spatial information needed for reconstruction. As a result, there is hardly a need for signal alignment. To search for an optimum setup, we first sweep each parameter used in dictionary learning, including signal dimension, batch size for training, regularization coefficient, and dictionary size. The derived parameters are then applied to the dictionary learning module. As the reconstructed signals are the linear composition of atoms in the trained dictionary, a ``clean" dictionary thereby have the denoising effect on signal reconstruction. To get a ``clean" dictionary, the training data is first filtered by a notch filter to remove power-line inference. Then the signal is passed through a band-pass filter to remove baseline wandering and high-frequency inference. Enabled by the pre-processing in the dictionary learning stage, the proposed framework eliminates the need of employing complicated pre-processing methods prior to random encoding on the sensor node. Instead, a simple segmentation module is sufficient for clean reconstruction. 

The initialization in dictionary learning is important. A poorly initialized dictionary may contain bad atoms that are never used \cite{mairalJ}. Generally, the dictionary can be initialized by random numbers or input data. For more difficult and regularized problem, it is preferable to start from a less regularized case and gradually increase the regularization coefficients. In our framework, the dictionary is initialized by randomly chosen columns from the input data set for simplicity.

The most notable advantage of ODL over other dictionary learning algorithms, such as K-SVD, is that ODL does not rely on the matrix factorization upon the entire training data. As a result, the time cost is much less compared to the non-online versions when handling large training datasets. So a specific input ECG signal that carries new features, such as disease information, can be quickly processed by the dictionary learning module to update the dictionary when necessary. As dictionary update does not depend on the previous samples, the framework also eliminates the demand of large storage space for prior inputs.

BP algorithm, running on the mobile node, is used in our framework to reconstruct high-quality signals. As ODL is compatible with other reconstruction algorithms, more computation efficient algorithms (e.g., fast iterative shrinkage-thresholding algorithm (FISTA) can be implemented to improve accuracy-complexity trade-off).

\begin{table}
	\caption{Performance Comparison of CS frameworks}
	\centering
	\begin{tabular}{ c c c } 
		\hline
		\hline
		Framework & CR & PRD ($\%$)\\ 
		\hline
		Proposed & 10 & 9 \\
		Ansari-Ram et al. \cite{6313763} & 5 & 9 \\ 
		
		Casson el al. \cite{6346966} & 4 & 9 \\ 
		
		Mamaghanian el al. \cite{5771056} & 3.4 & 9 \\
		
		Chae et al. \cite{6637862}& 2.5 & 9 \\ 
		\hline 
		\hline
	\end{tabular}
	\label{tab:partition}
\end{table}

\section{Experiment Results}
Experiments are conducted to compare the performance of the proposed framework in terms of recovery quality and CR with the conventional CS frameworks adopting pre-determined basis for the reconstruction of ECG signal. All frameworks employs the same random Bernoulli matrix $\mathbf{\Phi}$ (0/1 only) as the sensing matrix, so the hardware cost of the acquisition module, i.e., the sensor nodes, are the same. 

\subsection{Performance Metrics}
The compression ratio (CR) and percent root-mean-square difference (PRD) are used as the performance metrics.

1) Compression Ratio (CR):
CR is a measurement of the reduction of the data required to represent the original signal $\mathbf{f}$. If $m$ measurements are required to recover the signal with dimension $n$, then 
\begin{equation}
CR=\frac{n}{m}.
\end{equation}

2) Percent Root-mean-square Difference (PRD):
PRD is a measurement of the difference between the original signal $\mathbf{f}$ and the reconstructed signal $\mathbf{f'}$. As arbitrarily low PRD can be achieved by selecting a high DC level in signal $\mathbf{f}$, a more appropriate metric is to remove the DC bias in signal $\mathbf{f}$ as

\begin{equation}
	PRD=\frac{\Vert \mathbf{f-f'} \Vert_{2}}{\Vert \mathbf{f}-\mathbf{\bar{f}} \Vert_{2} } \times 100,
\end{equation}
where $\bar{\mathbf{f}}$ is the mean of signal $\mathbf{f}$.

\subsection{Experiment Settings and Results}
Through parameter sweeping, the dimension of the signal n is set to 256, size of the dictionary k is set to 512. Experiments are carried out based on the MIT-BIH Arrhythmia Database. In the experiments, 649984 samples are divided into 2539 epochs. Each epoch contains 256 samples. Among all the data sets, 512 epochs are randomly chosen to initialize the dictionary, 1621 epochs are used to train the dictionary, and the remaining is used as the testing set. For performance comparison, the pre-determined basis used in the reference framework is a joint basis composed by both discrete cosine transform (DCT) and descrete wavelet transform (DWT) bases \cite{7063062}. This is because the periods components (e.g. QS waves) and the spike components (e.g. R wave) have sparse representations on DCT and DWT basis, respectively. 

Figure 2 shows the performance comparison results. Overall, the proposed framework outperform the reference framework significantly due to the use of personlized basis in reconstruction . Specifically, an average PRD of 9$\%$, required for diagnosis purposes \cite{Zigel00theweighted}, can achieved at a high CR of 10x. This represents a 6.5x more sample size reduction (engergy saving) than the reference framework \cite{7063062}. Table 1 compares the proposed framework with existing CS frameworks \cite{5771056, 6313763, 6637862, 6346966} that adopt pre-determined basis in signal recovery. In general, our framework is able to further improve the CR by 2-4x for achieving an average PRD of 9$\%$. Fig.3 demonstrates the high reconstruction quality of the proposed framework in comparison to the reference framework \cite{7063062} when CR=10. 

\begin{figure}
	\centering
	\includegraphics[width=9cm]{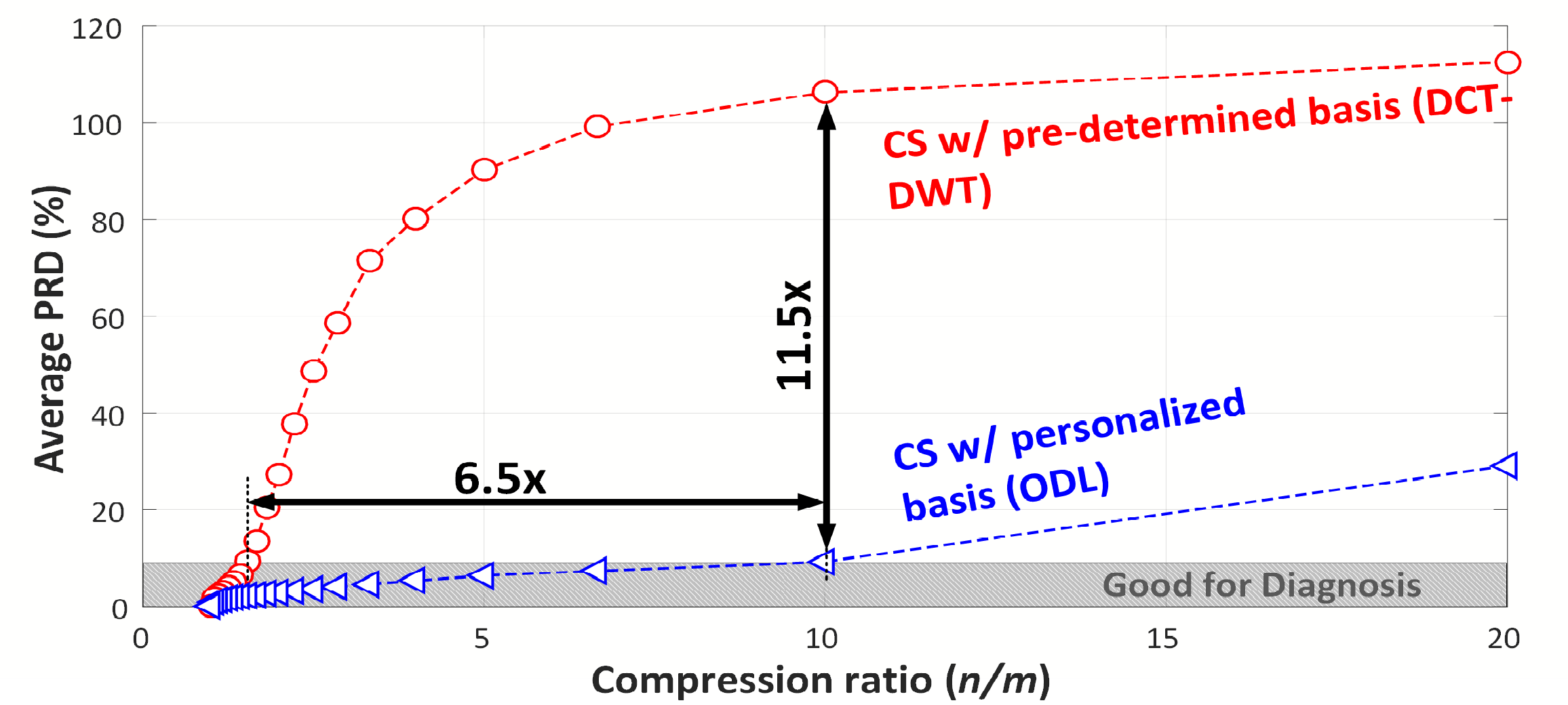}
	\caption{Comparison of our proposed framework with conventional CS framework in term of CR.}
\end{figure}

\begin{figure}
	\centering
	\includegraphics[width=9cm]{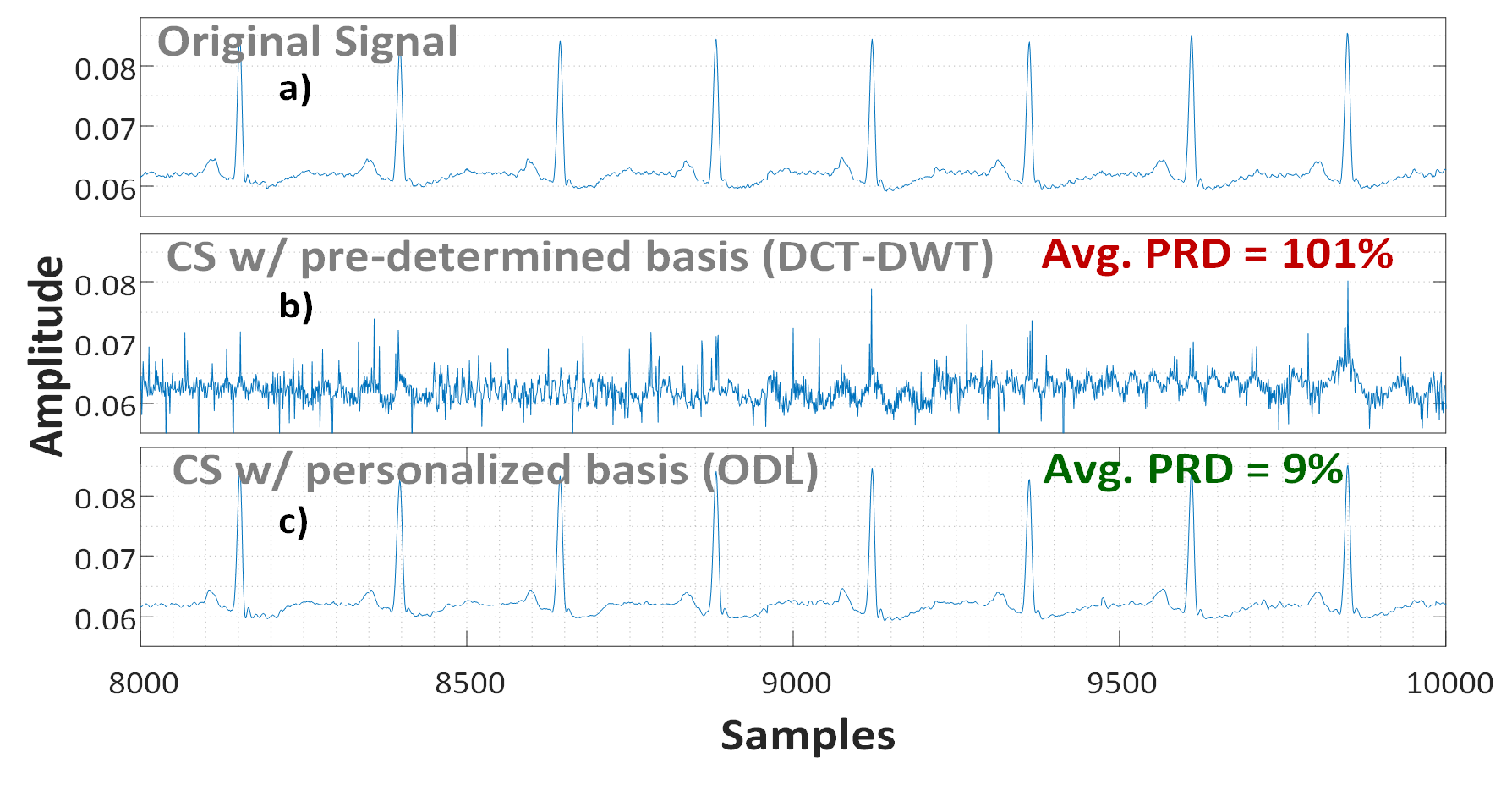}
	\caption{Reconstruction result for a segment of ECG signal when CR=10. (a) Original ECG signal; (b) Reconstructed signal using pre-determined DCT-DWT joint basis; (c) Reconstructed signal using online trained dictionary.}
\end{figure}

\section{Conclusions}
In this paper, we propose an energy-efficient data acquisition framework combining the notion of CS and ODL for long-term ECG monitoring. The framework significantly enhances CS performance by learning personalized basis to inform signal recovery. Experiment results show that by moving pre-processing to the dictionary learning stage, a simple segmentation process in the sensor nodes is sufficient to recover high-quality signals. In the future work, we will add sub-basis onto which the abnormal ECG signal is projected, when the ``healthy" sub-basis is unable to model the original signal accurately. 

\vfill\pagebreak
\clearpage
\bibliographystyle{IEEEbib}
\bibliography{references}

\begin{thebibliography}{10}

\bibitem{5771056}
H.~Mamaghanian et~al.,
\newblock ``Compressed sensing for real-time energy-efficient ecg compression
  on wireless body sensor nodes,''
\newblock {\em IEEE Trans. Biomed. Eng}, vol. 58, no. 9, pp. 2456--2466, Sep.
  2011.

\bibitem{Varshney}
U.~Varshney,
\newblock ``Pervasive healthcare and wireless health monitoring,''
\newblock {\em Mob. Netw. Appl.}, vol. 12, no. 2-3, pp. 113--127, Mar. 2007.

\bibitem{6155205}
F.~Chen et~al.,
\newblock ``Design and analysis of a hardware-efficient compressed sensing
  architecture for data compression in wireless sensors,''
\newblock {\em IEEE J. Solid-State Circuits}, vol. 47, no. 3, pp. 744--756,
  Mar. 2012.

\bibitem{islped_opt}
Y.~Wang et~al.,
\newblock ``Optimizing boolean embedding matrix for compressive sensing in rram
  crossbar,''
\newblock in {\em Proc. 2015 ACM/IEEE Int. Symp. Low Power Electron. and
  Design, Rome, Italy}, Jul. 2015, pp. 13--18.

\bibitem{6313763}
F.~Ansari-Ram et~al.,
\newblock ``Ecg signal compression using compressed sensing with nonuniform
  binary matrices,''
\newblock in {\em 16th CSI Int. Symp. on Artificial Intell. and Signal
  Process}, May 2012, pp. 305--309.

\bibitem{6637862}
D.H. Chae et~al.,
\newblock ``Performance study of compressive sampling for ecg signal
  compression in noisy and varying sparsity acquisition,''
\newblock in {\em IEEE Int. Conf. on Acoust., Speech and Signal Process}, May
  2013, pp. 1306--1309.

\bibitem{6346966}
A.J. Casson and E.~Rodriguez-Villegas,
\newblock ``Signal agnostic compressive sensing for body area networks:
  Comparison of signal reconstructions,''
\newblock in {\em Annu. Int. Conf. of the IEEE Eng. in Med. and Biol. Soc.},
  Aug. 2012, pp. 4497--4500.

\bibitem{CandesIntroCS}
E.J. Candes et~al.,
\newblock ``An introduction to compressive sampling,''
\newblock {\em Signal Process. Mag.}, vol. 25, pp. 21 -- 30, 2008.

\bibitem{1580791}
E.J. Candes et~al.,
\newblock ``Robust uncertainty principles: exact signal reconstruction from
  highly incomplete frequency information,''
\newblock {\em IEEE Trans. Inf. Theory}, vol. 52, no. 2, pp. 489--509, Feb.
  2006.

\bibitem{4011956}
M.~Elad and M.~Aharon,
\newblock ``Image denoising via sparse and redundant representations over
  learned dictionaries,''
\newblock {\em IEEE Trans. on Image Process.}, vol. 15, no. 12, pp. 3736--3745,
  Dec. 2006.

\bibitem{aharon}
M.~Aharon et~al.,
\newblock ``\rm k-svd: An algorithm for designing overcomplete dictionaries for
  sparse representation,''
\newblock {\em IEEE Trans. on Signal Process.}, vol. 54, no. 11, pp.
  4311--4322, Nov. 2006.

\bibitem{olshausen}
B.~A. Olshausen and D.~J. Field,
\newblock ``Sparse coding with an overcomplete basis set: A strategy employed
  by v1?,''
\newblock {\em Vision Res.}, vol. 37, pp. 3311--3325, Dec. 1997.

\bibitem{NIPS2006_2979}
L.~Honglak et~al.,
\newblock ``Efficient sparse coding algorithms,''
\newblock in {\em Advances in Neural Inform. Process. Syst.}, pp. 801--808. MIT
  Press, 2007.

\bibitem{Bottou08thetradeoffs}
L.~Bottou and O.~Bousquet,
\newblock ``The tradeoffs of large scale learning,''
\newblock in {\em Advances in Neural Inform. Process. Syst.}, 2008, pp.
  161--168.

\bibitem{mairalC}
J.~Mairal et~al.,
\newblock ``Online dictionary learning for sparse coding,''
\newblock in {\em Proc. of the 26th Ann. Int. Conf. on Mach. Learn.}, 2009, pp.
  689--696.

\bibitem{mairalJ}
J.~Mairal et~al.,
\newblock ``Online learning for matrix factorization and sparse coding,''
\newblock {\em J. Mach. Learn. Res.}, vol. 11, pp. 19--60, Mar. 2010.

\bibitem{7015915}
S.~Lee et~al.,
\newblock ``A new approach to compressing ecg signals with trained overcomplete
  dictionary,''
\newblock in {\em EAI 4th Int. Conf. on Wireless Mobile Commun. and
  Healthcare}, Nov. 2014, pp. 83--86.

\bibitem{5946515}
L.~F. Polania et~al.,
\newblock ``Compressed sensing based method for ecg compression,''
\newblock in {\em IEEE International Conference on Acoustics, Speech and Signal
  Processing (ICASSP)}, May 2011, pp. 761--764.

\bibitem{Abo}
M.~Abo-Zahhad et~al.,
\newblock ``Compression of ecg signal based on compressive sensing and the
  extraction of significant features,''
\newblock {\em Int. J. of Commun., Network and Syst. Sciences}, vol. 8, pp.
  97--117, 2015.

\bibitem{7063062}
F.~Ren and D.~Markovic,
\newblock ``18.5 a configurable 12-to-237ks/s 12.8mw sparse-approximation
  engine for mobile exg data aggregation,''
\newblock in {\em IEEE Int. Solid-State Circuits Conf.}, Feb. 2015, pp. 1--3.

\bibitem{Zigel00theweighted}
Y.~Zigel et~al.,
\newblock ``The weighted diagnostic distortion (wdd) measure for ecg signals
  compression,''
\newblock {\em IEEE Trans. Biomed. Eng.}, pp. 1422--1430, 2000.

\end{thebibliography}

\end{document}